# Domain wall dynamics driven by a localized injection of a spin-polarized current

G. Finocchio[1], N. Maugeri[2], L. Torres[3], B. Azzerboni[1]

[1] Department of Fisica della Materia e Ingegneria Elettronica, University of Messina, 98166 Messina, Italy.
[2] Department of Ingegneria Civile, University of Messina, 98166 Messina, Italy.
[3] Departamento de Fisica Aplicada, University of Salamanca, Plaza de la Merced s/n, 37008 Salamanca, Spain.

**This paper introduces an oscillator scheme based on the oscillations of magnetic domain walls due to spin-polarized currents, where the current is injected perpendicular to the sample plane in a localized part of a nanowire. Depending on the geometrical and physical characteristic of the system, we identify two different dynamical regimes (auto-oscillations) when an out-of-plane external field is applied. The first regime is characterized by nucleation of domain walls (DWs) below the current injection site and the propagation of those up to the end of the nanowire, we also found an oscillation frequency larger than 5GHz with a linear dependence on the applied current density. This simple system can be used as a tuneable steady-state domain wall oscillator. In the second dynamical regime, we observe the nucleation of two DWs which propagate back and forth in the nanowire with a sub-GHz oscillation frequency. The micromagnetic spectral mapping technique shows the spatial distribution of the output power is localized symmetrically in the nanowire. We suggest that this configuration can be used as micromagnetic transformer to decouple electrically two different circuits.**

## I. INTRODUCTION

The discovery that a spin-polarized current can apply a torque on the magnetization of a nanomagnet through transfer of spin angular momentum has opened a different way to manipulate magnetic states[1]. In particular, magnetization reversal and steady state dynamics have been observed in multilayer pillar elements (spin-valves and magnetic tunnel junctions)[2-4] and point contacts geometries[5]. Interactions with domain walls (DWs) have also been observed in nanowire, when a spin-polarized current is passed the flowing electrons exert a pressure on the DW that tends to move it in the direction of the electron flow[6, 7]. This current-induced motion of magnetic DWs is of interest for applications in magneto-electronic devices in which the DW is the logic gate or in the memory elements. If the magnetic nanowire contains DWs, an interaction mediated by the scattered charge carriers can give origin to a potential application of interacting DWs as a tunable nano-mechanical oscillator[8].

Recently, different systems based on the oscillations of magnetic domain walls due to spin-polarized currents have been studied[9-12], where for example the magnetization is pinned at the ends of a cylindrical ferromagnetic nanopillar[11], or a perpendicular material is used[12].

In a more complex system, three terminal spin-torque memory, a spin polarized current interacts in two different ways with the magnetization, first the current flows perpendicular to the sample plane into a localized part of the free layer, and then it flows in-plane up to the other contact[13]. The reverse of the free layer magnetization occurs via a nucleation of a DW below the injection site of the current and a consequently propagation of the DW across the free layer.

In this work, we study DW dynamics, when the current does not pass into the DW directly but it flows perpendicular to the nanowire plane and in a confined part of the nanowire itself. From experimental point of view, those systems can be realized using fabrication procedures and techniques similar to the ones used to build spin-valves with nano-aperture[14, 15]. Our results shows two different auto-oscillation mechanisms ((i) propagating and (ii) localized) of magnetization dynamics in which differently from other works are involved more than a DW: (i) nucleation of the DW below the injection site of the current and propagation up to the end of the nanowire; (ii) nucleation of two DWs which coupled move back and forth in the nanowire. Both configurations can be used as tuneable spin torque oscillators, the former in high frequency regime (larger than 5 GHz), the latter in sub-GHz regime. Furthermore in the latter regime, micromagnetic spectral mapping technique shows the power is localized symmetrically in the nanowire suggesting that this configuration can also be used as micromagnetic transformer.

## II. NUMERICAL DETAILS

We performed a numerical study of nanowires of size 1000nm x 40nm x 5nm and 500nm x 40nm x 5nm of Permalloy (Py) (negligible magneto-crystalline anisotropy, and near zero magnetostriction). Fig. 1 shows a sketch of the geometry under investigation and the Cartesian coordinate system we introduced. Differently by other works[9-12], the current flows perpendicular to the sample plane in a confined part of the nanowire and it does not pass into the DW directly, in particular the current is injected via a nanocontact ranging form 5x40nm$^2$ to 40x40nm$^2$ located as displayed in Fig.1. The structure of a DW (transverse or vortex) depends on a balance between exchange and anisotropy energies, for the geometrical and physical parameters we used in this work, it is possible to nucleate a transverse DW only[16].

The magnetization dynamics is studied by means of the numerical solution of the Landau-Lifshitz-Gilbert-Slonczweski equation with an effective field ($\mathbf{h}_{eff}$) which takes into account the magnetostatic, the exchange, the external, and the Oersted fields [17]:

$$\frac{d\mathbf{m}}{d\tau} = -(\mathbf{m}\times\mathbf{h}_{eff}) + \alpha\ \mathbf{m}\times\frac{d\mathbf{m}}{d\tau} + \tau(\mathbf{m},\mathbf{m_p}) \qquad (1)$$

where $\mathbf{m}$ is the normalized magnetization of the nanowire, $\alpha$ is the damping parameter and $d\tau$ is the dimensionless time. We consider a polarization $\mathbf{m_p}$ fixed along +x direction (1,0,0), and an expression for the torque[18]:

$$\tau(\mathbf{m},\mathbf{m_p}) = -\frac{g}{e\gamma_0}\frac{|\mu_B|j(x,y)}{M_s^2\ d}\varepsilon(\mathbf{m},\mathbf{m_p})\ \mathbf{m}\times(\mathbf{m}\times\mathbf{m_p}) \qquad (2)$$

where $g$, $\mu_B$, $\gamma_0$, and $e$ are the Landè factor, the Bohr magneton, the gyromagnetic ratio and the electron charge respectively, $M_s$, and d are the saturation magnetization and the thickness of the nanowire, $j(x,y)$ is the current density distribution which is uniform below the injection site and zero outside[19]. The $\varepsilon(\mathbf{m},\mathbf{m_p})$ is the polarization function:

$$\varepsilon(\mathbf{m},\mathbf{m_p}) = 0.5P\Lambda^2 /\left(1+\Lambda^2 + (1-\Lambda^2)\mathbf{m}\bullet\mathbf{m_p}\right) \qquad (3)$$

being $P$ and $\Lambda^2$ torque parameters[18]. We use the following simulation parameters: exchange constant $1.3\ 10^{-11}$ J/m, damping 0.02, $M_s = 650\ 10^3$ A/m, $d$=5 nm, $P$=0.38 and $\Lambda^2$=2.5.

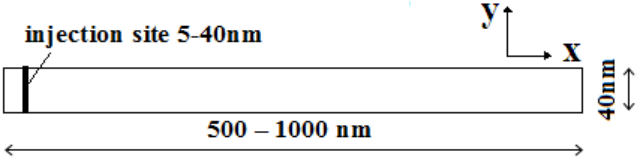

Fig. 1.Sketch of the device geometry studied in this numerical experiment.

## III. RESULTS

We performed a systematic study of the magnetization dynamics driven by a localized spin-polarized current (the current density ranging from 1 to $5\ 10^8$ A/cm$^2$) and an out-of-plane external field of 500mT, 750mT and 1000mT. For most configurations the magnetization dynamics is characterized by chaotic behaviour with inhomogeneous spatial distribution of magnetization. Our numerical results also show coherent magnetization dynamics, in particular we observe two different regimes: propagating and localized. The former regime is observed in 1000nm x 40nm x 5nm nanowire. Fig. 2 shows the oscillation frequency as function of the current density ($H$=750mT, size of injection site 40x40nm$^2$) computed by means of the micromagnetic spectral mapping technique (a typical power spectrum, computed for $J$=2.5 $10^8$ A/cm$^2$, is displayed in the inset of Fig. 2)[20-22]. The quasi-linear dependence is characterized by a slope of 1.8 (GHz cm$^2$/A), and the "blue shift" functional dependence is characteristic of the out-of-plane precession[23], as can be observed. The mechanism related to that self-oscillation dynamics can be understood by the analysis of the spatial distribution of the magnetization. Fig. 2 (bottom) shows some characteristic snapshots of the magnetization, below the injection site of the current the magnetization continuously is reversed, rotating from the alignment along +x (red) to the alignment along –x (blue) (snapshots A and C), giving rise to a nucleation of a transverse DW in the boundary of the injection site (left part). The DW separates configuration of magnetization with 180° domains. The domain nucleated is expelled at the end of the nanowire via a propagation mechanism. Similar results but for smaller range of current are observed for different values of external fields. An advantage of this configuration with respect to the one presented in Ref[12] is the larger frequency of the magnetization dynamics, it is tuneable in a wide range of frequency with a quasi-linear behaviour as function of current, in line of principle it is possible to increase the output power by means of the synchronization the effect of several contacts localized in different regions of the nanowire[24, 25], but in contrast the current density involved are larger and it is necessary to apply an external field.

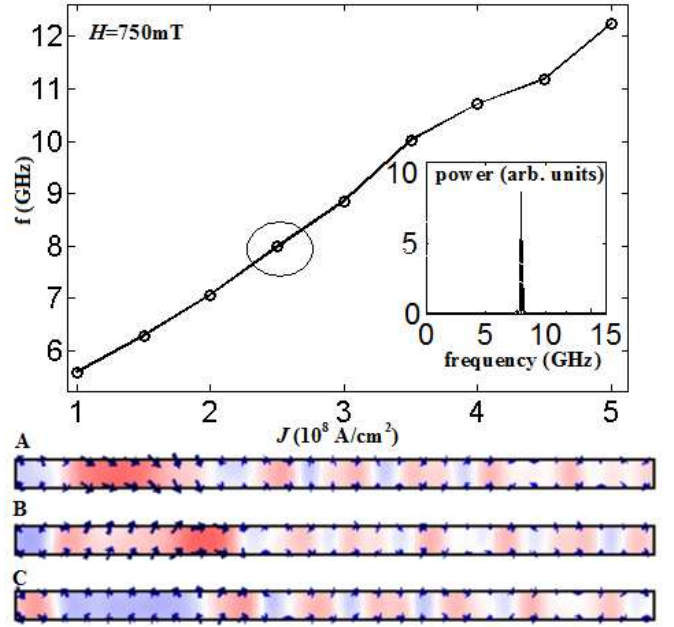

Fig. 2. Main panel: functional dependence of the oscillation frequency of the magnetization dynamics as function of the current density (injection site 40x40 nm$^2$) in a nanowire of 1000nm x 40nm x 5nm, $H$=750mT. The inset shows an example of power spectrum computed via the micromagnetic spectral mapping technique for a $J$=2.5 $10^8$ A/cm$^2$. The bottom part of the figure shows snapshots (red and blue mean positive and negative component of the magnetization along the x axis, the arrow indicate the direction of the in-plane component of the magnetization) of the spatial distribution of the magnetization which characterizes the magnetization dynamics in the propagating regime.

Decreasing the size of the nanowire and for a smaller dimension of the injection site for the current (5 nm x 40nm), we find out a different dynamical regime (localized). For a 500nm x 40nm x 5nm nanowire, an out-of-plane field of 500mT, in a range of current from $J$=0.4 $10^8$ A/cm$^2$ to $J$=0.6 $10^8$ A/cm$^2$, we observe the nucleation of two DWs which coupled move back and forth in the nanowire giving rise to a microwave signal with a main oscillation frequency around 800MHz (the frequency is almost independent on the current density). Basically, the energy pumped by the spin current is enough to compensate the energy dissipated by the DWs motion due to the damping parameter. A zoom of the time domain trace of the x-component of the average magnetization is displayed in the main panel of Fig. 3 for $J$=0.5 $10^8$ A/cm$^2$. Fig. 3 (bottom part) also shows the spatial distribution of the magnetization in the points A, B, C, D, and E. The nanowire is characterized by an alternating of 360° domain configuration (A, C, E) with 180° domain configuration (B, D). As can be observed from the power spectrum of Fig. 4 computed for a current density $J$=0.5 $10^8$ A/cm$^2$, the magnetization dynamics is characterized by few harmonics. Our numerical results show that increasing the size of the injection site (larger than 10 nm x 40nm) and for the current we simulated the stationary dynamics is not present, in particular chaotic magnetization dynamics is observed[26, 27].

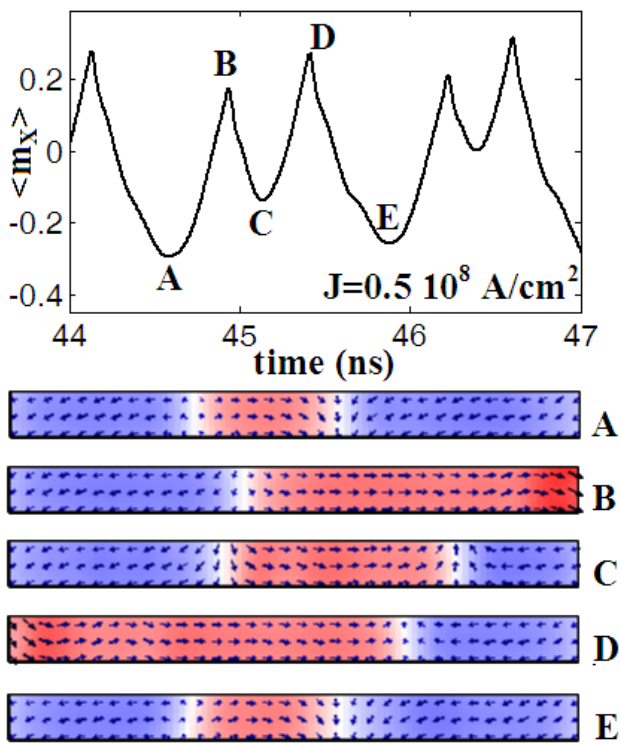

Fig. 3. Time trace of the x-component of the average x-component of the magnetization computed for $J$=0.5 $10^8$ A/cm$^2$ and $H$=500mT. The bottom part of the figure shows snapshots (red and blue mean positive and negative component of the magnetization along the x axis, the arrow indicate the direction of the in-plane component of the magnetization) of the spatial distribution of the magnetization which characterizes the magnetization dynamics in the localized, the points A, B, C, D, E are the same of the time trace.

We use the micromagnetic spectral mapping technique to determine the spatial distribution of the modes as shown in the insets of Fig. 4 (power increases from white to black). One interesting result is the symmetric spatial distribution of the oscillation power. Our results suggest the possibility to use this system to decouple the dynamics in two well separated devices (four terminals device), in other words it is possible to excite dynamics using a spin valve located in a side of the nanowire to obtain a microwave signal in a different position of the nanowire itself, this signal can be read, for example, by introducing a magnetic tunnel junction (MTJ). The idea is similar to the three terminal device presented in Ref[13], but with the spin valve and the magnetic tunnel junction separated electrically. The behaviour of this system can be seen as a nanoscale micromagnetic transformer where the energy available in the secondary winding (MTJ) depends on the tunnelling magneto-resistance ratio.

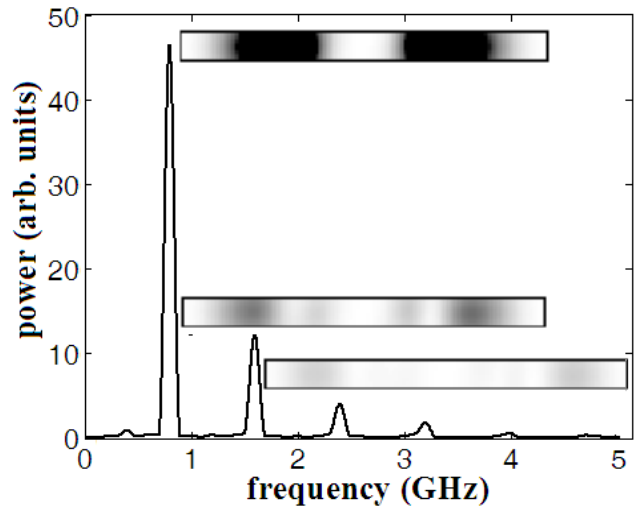

Figure 4: Power spectrum of the magnetization dynamics for $J$=0.5 $10^8$ A/cm$^2$ and $H$=500mT computed by means of the micromagnetic spectral mapping technique. The insets are the spatial distribution of the output power of the excited modes (the power increases from white to black)

## IV. CONCLUSIONS

We performed a systematic study of the magnetization dynamics driven by a spin-polarized current injected in a localized part of a magnetic nanowire. Depending on the geometrical and physical characteristic of the nanowire, we identify two different dynamical regimes (auto-oscillations) (i) propagating and (ii) localized. In both, regimes are involved more than a DW: (i) nucleation of the DW below the injection site of the current and propagation up to the end of the nanowire; (ii) nucleation of two DWs which coupled move back and forth in the nanowire. In the propagating regime, this system can be used as a wide range tuneable DW oscillator. In the localized regime, the output power is localized symmetrically and in a reduced part of the nanowire, our results suggest the possibility to use this system as micromagnetic transformer. Furthermore, in those geometries, a recent experiment shows high DW velocity excited at significantly lower current densities than in nanowires with current applied along the nanowire[28].

ACKNOWLEDGMENT

This work was partially supported by Spanish Project under Contracts No. MAT2008-04706/NAN and No. SA025A08.